\documentclass[sigplan,screen, nonacm]{acmart}

\AtBeginDocument{%
  \providecommand\BibTeX{{%
    \normalfont B\kern-0.5em{\scshape i\kern-0.25em b}\kern-0.8em\TeX}}}
    
\setcopyright{none}
\begin{document}

\title[Enhancing Empathy in Virtual Reality]{Enhancing Empathy in Virtual Reality: An Embodied Approach to Mindset Modulation}

\author{Seoyeon Bae}
 \affiliation{%
  \institution{Seoul National University}
  \city{Seoul}
  \country{Republic of Korea}}
 \email{cathy4339@snu.ac.kr}
 \orcid{0000-0002-8323-6996}

\author{Yoon Kyung Lee}
 \affiliation{%
  \institution{Seoul National University}
  \city{Seoul}
  \country{Republic of Korea}}
 \email{yoonlee78@snu.ac.kr}
 \orcid{0000-0002-5577-6311}
  
\author{Jungcheol Lee}
 \affiliation{%
  \institution{Joongbu University-Goyang}
  \city{Goyang}
  \country{Republic of Korea}}
 \email{jclee.cjf@gmail.com}
 \orcid{0009-0001-6009-1069}

\author{Jaeheon Kim}
 \affiliation{%
  \institution{Joongbu University-Goyang}
  \city{Goyang}
  \country{Republic of Korea}}
 \email{fldh753@jmail.ac.kr}
 \orcid{0009-0004-7486-7528}

\author{Haeseong Jeon}
 \affiliation{%
  \institution{Joongbu University-Goyang}
  \city{Goyang}
  \country{Republic of Korea}}
 \email{seng192@jmail.ac.kr}
 \orcid{0009-0000-6250-1917}

\author{Seung-Hwan Lim}
 \affiliation{%
  \institution{Joongbu University-Goyang}
  \city{Goyang}
  \country{Republic of Korea}}
 \email{come112008@jmail.ac.kr}
 \orcid{0009-0007-1796-2256}

\author{Byung-Cheol Kim}
 \affiliation{%
  \institution{Joongbu University-Goyang}
  \city{Goyang}
  \country{Republic of Korea}}
 \email{ciel@jbm.ac.kr}
 \orcid{0009-0008-0837-4716}

\author{Sowon Hahn}
 \affiliation{%
  \institution{Seoul National University}
  \city{Seoul}
  \country{Republic of Korea}}
 \email{swhahn@snu.ac.kr}
 \orcid{0000-0002-2533-4002}


\renewcommand{\shortauthors}{Bae et al.}

\begin{abstract}

A growth mindset has shown promising outcomes for increasing empathy ability. However, stimulating a growth mindset in VR-based empathy interventions is under-explored. In the present study, we implemented prosocial VR content, ``\textit{Our Neighbor Hero}\footnote{Our demo video can be viewed here: \href{https://youtu.be/8tg6ydFjPyE}{https://youtu.be/8tg6ydFjPyE}},'' focusing on embodying a virtual character to modulate players' mindsets. The virtual body served as a stepping stone, enabling players to identify with the character and cultivate a growth mindset as they followed mission instructions. We considered several implementation factors to assist players in positioning within the VR experience, including positive feedback, content difficulty, background lighting, and multimodal feedback. We conducted an experiment to investigate the intervention's effectiveness in increasing empathy. Our findings revealed that the VR content and mindset training encouraged participants to improve their growth mindsets and empathic motives. This VR content was developed for college students to enhance their empathy and teamwork skills. It has the potential to improve collaboration in organizational and community environments. 

\end{abstract}


\begin{teaserfigure}
  \includegraphics[width=\textwidth]{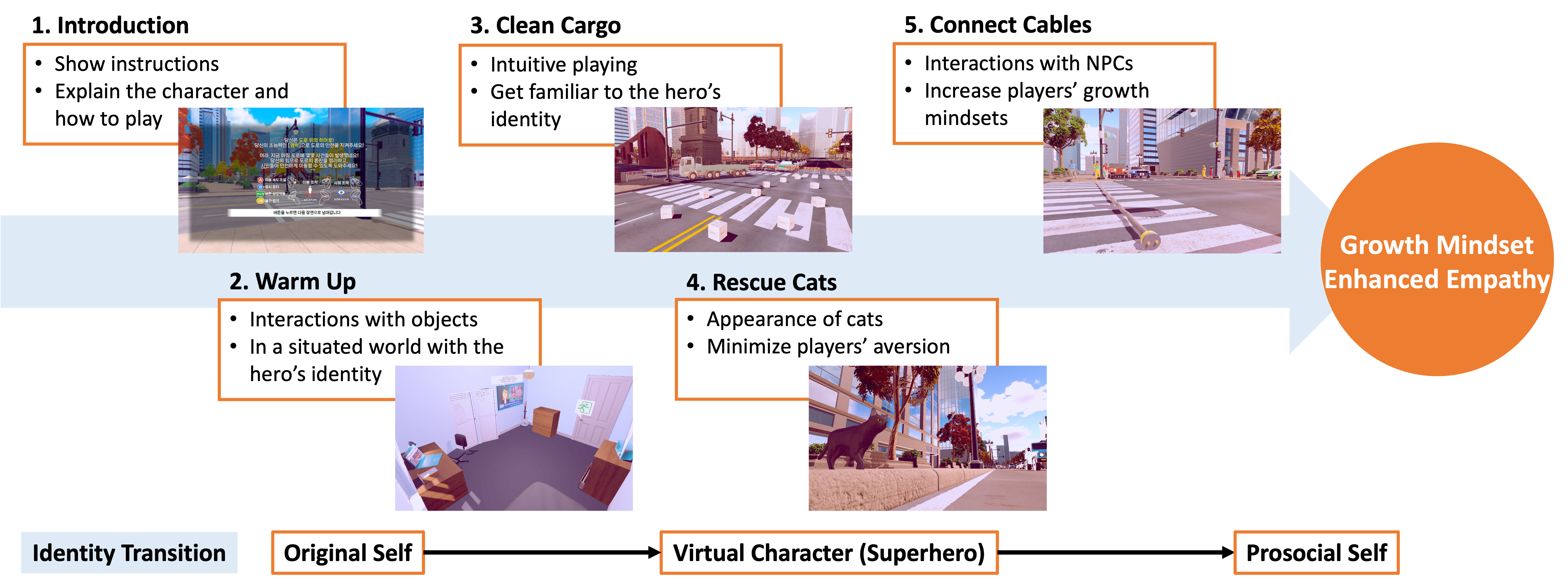}
  \caption{The procedure of \textit{Our Neighbor Hero}. The VR content includes prosocial activities and transforms players' identities.}
  \Description{The figure explains how to stimulate players' growth mindsets and empathy. The VR content includes five steps: introduction, warming up, cleaning cargo, rescuing cats, and connecting cables. By experiencing the content, players modified their identities from the original to prosocial ones.}
  \label{fig:teaser}
\end{teaserfigure}

\maketitle

\section{Introduction}
As social beings, humans benefit from the ability to empathize with and understand others, which plays a crucial role in fostering cooperation and resolving conflicts throughout their lives. Empathy is understanding and feeling other peoples' emotions \cite{davis1980multidimensional, davis1983measuring}. This ability is essential for people to engage in social activities and build relationships. A high empathy ability helps people collaborate in workplaces, resolve conflict, and foster deeper interpersonal relationships \cite{kelly2001mood, hallowell1999human}. 

In the past, empathy was considered an unchanging trait \cite{eysenck1978impulsiveness, davis1983measuring}. More alternative views of empathy are that it is a malleable skill that can be developed and improved through regular practice, similar to a ‘muscle’ \cite{zaki2019war}. This approach has led to increased attention to the potential of ‘empathy interventions’, training methods that encourage individuals to practice empathic skills such as perspective-taking and emotion-sharing on a regular basis. These empathy interventions have shown promising results in enhancing one’s empathic ability \cite{weisz2022brief}. 

Empathy interventions have become increasingly important in various domains, such as training healthcare practitioners, where understanding patients’ or clients’ emotions is pivotal. Research shows that training empathy effectively builds rapport between patients and medical practitioners \cite{kelm2014interventions}, increasing compassion towards patients \cite{hofmeyer2021strategies}. A recent trend in empathy intervention includes promoting a \textit{growth mindset} — a belief that basic skills can improve with effort \cite{dweck1988social, yeager2012mindsets, paunesku2015mind, yeager2019national}. Having a growth mindset regarding empathy means believing one can enhance or improve one's empathy ability through motivation and effort \cite{schumann2014addressing, weisz2021building}. Training programs to reinforce growth mindsets include watching video stimuli or writing letters \cite{schumann2014addressing, weisz2021building, weisz2022brief}. 

Virtual Reality (VR) has recently been known as the ``ultimate empathy machine'' \cite{bailenson2018experience, milk2015, zaki2019war}. VR helps players better grasp others' perspectives as if they were ``walking in their shoes'' \cite{herrera2018building, ho2022perspective, van2018virtual}. Players immerse themselves in unique experiences they would not normally encounter. For example, they might live a day in the life of a homeless person \cite{herrera2018building} or spend time inside a refugee camp \cite{schutte2017facilitating}. These unique experiences elicit motivation to help, leading to increased prosocial behaviors (e.g., signing a petition for policy to help homeless people)  \cite{herrera2018building}.

Immersive VR experiences bridge players and the virtual character, facilitating players to understand different perspectives more deeply. This synchronization of distinct identities is called \textit{character identification}, a core mechanism of training empathy through VR  \cite{gupta2020investigating}. 

However, empathy interventions targeting players' mindsets using VR content remain under-explored. We expect that exposing players to unfamiliar situations will stimulate growth mindsets and motivation to help those in need. 

The present study implemented VR content for empathy training, ``\textit{Our Neighbor Hero},'' to prompt players' growth mindsets and encourage prosocial behaviors. We conducted an experimental study to investigate whether our VR content is effective for empathy training, with the following research question: \textbf{\textit{How do we implement prosocial VR content to enhance one's growth mindset and empathy?}}

We, therefore, incorporated the superhero concept as the central character, creating a natural context for players to practice prosocial behaviors within the virtual environment. Our goal was to encourage players to assist others within the VR content while identifying the original self with the virtual character, reinforcing a growth mindset. We also considered various details in the content, including positive feedback, the content's difficulty, light sources and background colors, animations, interactive elements, and sensory feedback.

\section{Related Work}
\subsection{Empathy Intervention and Growth Mindset}
Empathy is a multifaceted concept, including cognitive and affective aspects \cite{neff2003self, de2006vignemont}. Cognitive empathy indicates understanding others’ emotions and perspectives \cite{eisenberg2014altruistic}. Affective empathy represents an ability to feel others' emotions simultaneously, which is found effective in prosocial decisions like vaccination during the COVID-19 pandemic \cite{pfattheicher2022information} and charity donations \cite{tusche2016decoding}. The motivational aspect is also a part of empathy, including the desire to relieve others’ emotional pain \cite{zaki2019war}. 

Continuous efforts have been made to strengthen empathy through training \cite{eisenberg1987relation, lam2011empathy, zaki2019war}. Empathy interventions have primarily focused on technique-based methods, including perspective-taking, recognizing and responding to others' emotions, and experiencing interpersonal skills \cite{batt2013teaching, teding2016efficacy}. However, whether these skill-based interventions were sufficient was raised because they depended on context or one's psychological state \cite{lam2011empathy, weisz2021building}. One way to overcome the limitations of former empathy interventions was using a growth mindset \cite{schumann2014addressing, weisz2021building}.

\textit{Mindsets} are beliefs about the nature of specific characteristics people possess, such as intelligence or personality \cite{dweck2006mindset, dweck2012mindsets, yeager2012mindsets}. Mindsets have two subtypes: growth mindset (\textit{incremental theory}) and fixed mindset (\textit{entity theory}) \cite{dweck1988social, dweck1999self}. When people endorse a growth mindset, they try to improve their ability based on setbacks and face challenges. On the other hand, those who have a fixed mindset believe that one's traits or attributes are complex to change and thus try to validate their ability by avoiding challenges.

Mindsets play a critical role in empathy intervention. The motivational aspect of empathy can be influenced by an individual's mindset and societal norms \cite{schumann2014addressing, weisz2021building, weisz2022brief}. \cite{schumann2014addressing} found that people who believed they could improve their empathy skills through effort were likely to show higher empathic effort and willingness to help others than those who did not think so. Similarly, a recent study by \cite{weisz2021building} found that interventions targeting a growth mindset improved the empathy ability of first-year undergrad students.

\subsection{Virtual Reality for Enhancing Empathy}
VR experiences maximize the effects of empathy training by providing players with immersive experiences through multimodal channels, such as visual, auditory, and haptic ones \cite{shams2008benefits, sigrist2013augmented}. Players also treat virtual objects as if they were real, even though they are aware that the environment they are playing in is not real \cite{sanchez2005presence}. \textit{Psychological Presence} is the feeling of ``being there'' \cite{bailenson2018experience}; in other words, the perception of being surrounded by a virtual environment \cite{weber2021get}. 

The \textit{General Learning Model} explains how VR content directly affects players' empathy and prosocial behaviors \cite{buckley2006theoretical, gentile2009effects}. Experiences in virtual environments shape players' internal state, and the internal state determines which types of behaviors to choose in the physical world. If players experience empathizing with others and helping them in VR content, they will likely help others when they face similar situations in physical reality. 

Identification with a virtual character and identity transition is critical in prompting empathy through VR experiences \cite{gupta2020investigating}. Players think of their virtual bodies as an extension of the self in virtual environments \cite{bailey2016does}. Players embrace virtual characters' traits and transform their self-concepts by identifying with the characters and taking their perspectives \cite{klimmt2009video}. They then comply with behaviors satisfying the standards of their changed self-concepts \cite{burke2006identity}.

Virtual Reality Perspective-Taking (VRPT) has been shown to be an effective method for increasing empathy and promoting prosocial behaviors \cite{herrera2018building, ho2022perspective, schutte2017facilitating}. The 3D VR video of a girl living in a refugee camp, ``\textit{Clouds Over Sidra},'' made participants feel close to and empathize with refugees by vividly experiencing her daily life \cite{schutte2017facilitating}. Becoming homeless and getting through their life in VR content also helped players comprehend their states, finally leading to practical behaviors such as signing a petition for people without homes \cite{herrera2018building}. One study created a VR game for saving robots by extinguishing fire with water guns \cite{ho2022perspective}. This study inspired participants to take perspectives of non-player characters (NPCs) by protecting them from fire. VRPT influenced their perceived closeness to the NPCs, and the closeness affected participants' empathy towards the NPCs.

\begin{table*}[!htpb]
  \caption{Technical details considered in designing \textit{Our Neighbor Hero}}
  \label{tab:cons}
  \resizebox{\textwidth}{!}{%
  \begin{tabular}{cl}
    \toprule
    Considerations& Details \\
    \midrule
    Light source placement & \begin{tabular}[c]{@{}l@{}}Controlled the contrast of objects using rendering shader function or depth textures \\ Used directional light for outdoor scenes, while used point light for indoor scenes \end{tabular} \\ \hline
    Background color grading & Adjusted colors of the background, including skybox, trees, and buildings, vividly and naturally \\ \hline
    Animations & \begin{tabular}[c]{@{}l@{}}Fine-tuned the speed of boxes, cats, cars, and cables \\ Made objects' movements smooth by diversifying possible motions\\\end{tabular} \\ \hline
    Interactive user-behaviors & \begin{tabular}[c]{@{}l@{}}Players grabbed objects from a distance with superpowers \\ Players controlled their movements and points of view with left- and right-hand controllers \\ Players used the triggers and the grip buttons to click and catch objects \end{tabular}\\ \hline
    Guidance of interactive behaviors & Included guiding functions of how to play, such as pointers, colors, and transparent sockets\\ \hline
    Auditory and haptic feedback & \begin{tabular}[c]{@{}l@{}}Contained auditory responses, including urban noises and sounds of people cheering and cats meowing \\ Added haptic responses to guide players' behaviors
    \end{tabular}\\ \hline
    Textual feedback & \begin{tabular}[c]{@{}l@{}} Conveyed immediate feedback regarding players' behaviors \\ Considered contents of the feedback appropriate for the situations \end{tabular} \\
  \bottomrule
\end{tabular}}
\end{table*}

\section{Design}
\subsection{Scenario}
We developed a prosocial VR experience program and named it \textit{Our Neighbor Hero}. The program comprised an introduction, warm-up in the superhero's room, and three missions – cleaning spilled cargo, rescuing cats trying to cross the road, and connecting cables to traffic lights.

\subsection{Mindsets-related Considerations}
\subsubsection{Virtual Character Setting}
We adopted a friendly superhero who helps his neighbors daily to change players' mindsets. We intended for players to augment their thoughts that ``we can help others'' through the identification process \cite{bailey2016does, gupta2020investigating, klimmt2009video}. We predicted that identifying with the superhero induced players to alter their mindsets and behave prosocially. Therefore, we set the virtual character as a bridge between players' original and altruistic identities and let players' identities transform into prosocial ones.

Our hero acts like one of many average characters who commute to work or school, cross the roads, and shop for groceries. At the same time, he has superpowers and helps others in emergencies. We designed for players to feel comfortable around the virtual neighborhood and being the character so they could identify with the hero and endorse growth mindsets. Emotional closeness is crucial for empathizing with a target \cite{beeney2011feel, paiva2005learning}. Conversely, unrealistic characters in content hinder players' identity shifts \cite{klimmt2009video}. To support the identification process, we included backgrounds and objects likely found in real life, such as infrastructures of many urban metropolises (e.g., roads, buildings) and a college student dorm-like room.

As a result, we planned for players to cultivate a growth mindset by identifying with the superhero. The virtual character served as a stepping stone towards a prosocial identity, making players continue altruistic behaviors in the physical world. 

\subsubsection{The Content Sequences}
Figure \ref{fig:teaser} shows the flow of the content. The sequences gradually stimulated players' growth mindsets and made them feel they were the main characters. Consequently, we included instructions at the start phase, explaining who the main character was and how players could manipulate the controllers (1. Introduction in Figure\ref{fig:teaser}).

After the introduction, players moved to the Warm-up phase (2. Warm-up in Figure \ref{fig:teaser}). In this phase, players entered the superhero room. The room made players identify with the character before they started the missions. Daily objects in the room were designed to be interactive so players could feel they were in the virtual world. Three electronic devices (a television, a laptop, a radio) were placed in the room, and players were transported to the three missions (\textit{Clean Cargo}, \textit{Rescue Cats}, \textit{Connect Cables}), as each electronic device was connected to each mission. Players were briefed with news of nearby accidents through the electronic devices, and selecting each device allowed them to teleport to the location. After completing each mission, players returned to the room and went to the next step. This composition raised the scenario's stability and prompted players to cultivate their growth mindsets. 

The order of the three missions was adjusted to alter players' mindsets step by step. We arranged the \textit{Clean Cargo} mission as the first, so players became accustomed to the superhero and his identity (3. Clean Cargo in Figure \ref{fig:teaser}). Since grabbing a box and putting it on the truck was intuitive, players knew how the content worked. We then put the \textit{Rescue Cat} mission in the middle of the content to prevent players from feeling resistance to abruptly changed mindsets (4. Rescue Cats in Figure \ref{fig:teaser}). This organization refreshed the content's overall atmosphere. Finally, we placed the third mission, \textit{Connect Cable}, as the last (5. Connect Cables in Figure \ref{fig:teaser}). Since the third one provided players with more interactions with NPCs (citizens) than any other two missions, we predicted this was appropriate for the final step to boost growth mindsets.

\subsection{Implementational Considerations}

We also considered additional factors like positive feedback, content difficulty, and duration to reinforce the effects of growth mindsets. Minor details about implementational considerations are also illustrated in Table \ref{tab:cons}.

\subsubsection{Positive Feedback with Interactions}
Positive feedback was offered to players in the content so that they directly recognized the results of their behaviors. Positive feedback, like expressing gratitude to helpers, is a powerful motivator for prosocial behaviors \cite{grant2010little, guo2017influence, mccullough2001gratitude}. We, therefore, encouraged players to feel joy about their behaviors by inserting scenes of positive feedback through interactions with NPCs.

For example, the truck driver appreciated for helping when the players accomplished the first mission. The rescued cats and citizens on the road cheered them up when players approached them while carrying out the tasks. They also expressed gratitude to players after the problems had been solved. 

Furthermore, we presented the status of the progress (i.e., scores) on the top right of the screen. This scoreboard was intended for players to feel a sense of accomplishment and get the motivation to fulfill their missions. The score box instantly changed when they succeeded in cleaning up cargo, saving cats, or connecting cables.

\subsubsection{Difficulty and Duration}
Each mission had a time limit of five and a half minutes to control the content's difficulty. We encouraged players to feel accomplished while maintaining tension. Making games or VR content challenging to complete drives players to become deeply immersed in games \cite{lieberman2006can}. Thus, setting an appropriate duration to maintain players' interest without failing missions was crucial. We did a series of beta tests and concluded that five and a half minutes were proper for both skilled and novice players while trying to finish in a given time. 

The time limit was presented on the top left of the screen so players could check the time left without being distracted. The time was initially displayed with a white clock icon in a black box but switched into a red fire if players ran out of time.

\subsection{Apparatus}
We created the VR content with the VR game engines Unity (ver. 2021.3.8.f1), Visual Studio 2019, and Blender. We tested it in different settings to enhance compatibility with our content. During the development, we initially tested with Dell Precision 7910 Tower workstation (CPU $=$ Intel(R) Xeon E5-2640 v4 with 2.4GHz speed, RAM $=$ 64GB, graphic card $=$ NVIDIA Quadro M5000).

In the experiment, we used Meta Quest 2. The size of the machine is 8.8 $*$ 17.7 inches, with 503g of weight. The machine has a 128GB storage capacity with 16GB RAM. It tracks users' movements with 6 degrees of freedom and lets them view virtual environments at a resolution of 1832 $*$ 1920 pixels per eye. Players can interact with the environments via two-hand controllers. They do not need headphones to listen to the sound as 3D positional audio is built into the headset, and glasses are compatible. 

The laptop storing the VR content and casting the machine's screen was ASUS ROG Zephyrus G15. We used a laptop to run the program during the experiment with specific specs (CPU $=$ AMD Ryzen 7 6800HS with Radeon Graphics 3.20 GHz, RAM $=$ 16GB, SSD $=$ 512 GB, graphic card $=$ NVIDIA GeForce RTX 3060). 

\section{Experiment}
A total of 54 college students participated in the study to examine the effect of the VR content and the mindset training. They were recruited through the participant recruitment system and the student communities at Seoul National University ($N_{male}$ = 19, $M_{age}$ = 21.43, $SD_{age}$ = 2.54). 

We conducted the study with a between-subject design. Participants of the first group (i.e., the ``VR-only'' group) only experienced the VR content. The second group (the ``combined'' group) played the  VR content and received training on growth mindset by reading a short research paper about the malleability of empathy \cite{schumann2014addressing, weisz2021building}. We translated and revised its original version \cite{schumann2014addressing, weisz2021building} into Korean. It contained information and experimental cases in which individuals promoted empathy by believing their empathy could increase through effort. Participants were asked to remind themselves of the article's details and what they learned from reading it. By doing so, participants were able to adopt a growth mindset on their capabilities for growing empathy. Both the combined and the VR-only groups comprised 27 individuals each, and participants were randomly assigned to the two groups.

After learning how to play, all participants entered the virtual environment and started to experience the VR content. The researchers kept checking whether they felt sick during the experiment and let them take a rest if they reported symptoms. It took about 10 minutes to complete. Next, the combined group read the research paper on a growth mindset, while the VR-only group moved directly to the survey stage. Participants' empathic motives, empathy, and prosocial behavior levels were measured in this stage, including demographic information such as age, gender, education level, and income. At the end of the experiment, participants provided additional feedback on the VR content and their impression of the experience in general. This study was approved by the Institutional Review Board of Seoul National University (IRB No. 2207/002-010).

\section{Results}
The data and code for analyses are available\footnote{ \href{https://github.com/Seoyeon-Bae/Our-Neighbor-Hero-VR.git}{https://github.com/Seoyeon-Bae/Our-Neighbor-Hero-VR.git}}.

\subsection{Empathy Ability Questionnaires}
\subsubsection{Normality Tests} We first tested whether the data followed the normal distribution. As the sample size of both groups was small (\textit{N} < 30), we checked quantile-quantile (Q-Q) plots and distribution histograms. We also conducted the Shapiro-Wilk normality test \cite{shapiro1965analysis}. According to the tests, the hypothetical prosocial behavior variable (\textit{W} = .83, \textit{p} < .001) did not follow normal distributions. Therefore, we analyzed scores for the variable using the non-parametric Mann-Whitney U test \cite{mann1947test}.

\begin{figure}[htpb]
    \begin{center}
    \includegraphics[width=\columnwidth]{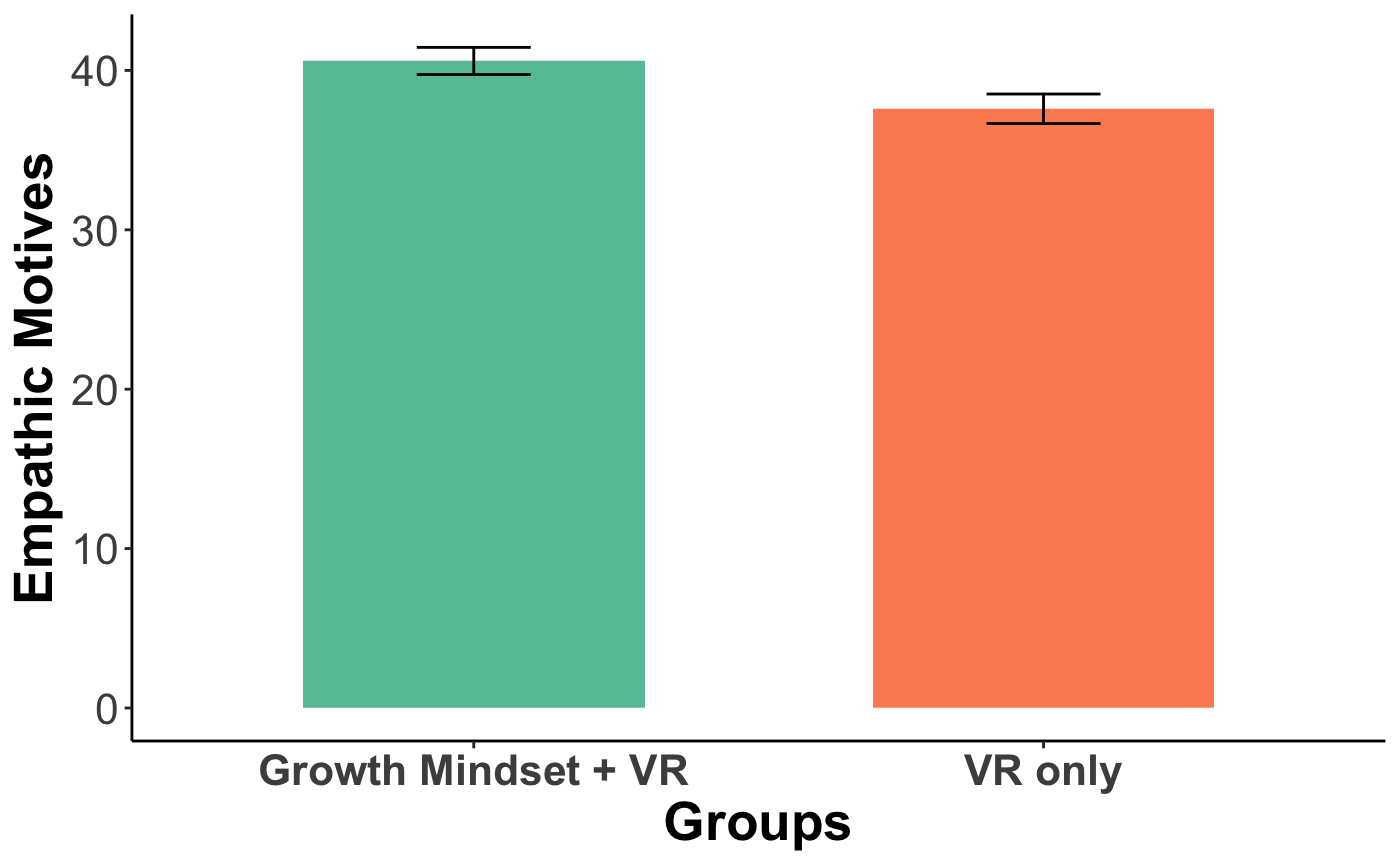}
    \end{center}
    \caption{Mean empathic motives ratings between participants who received growth mindset training and VR experience and those experienced only VR.}
    \label{fig:2}
\end{figure}

\subsubsection{Empathic Motives} Participants responded to the Empathic Motives Questionnaire \cite{schumann2014addressing, weisz2022brief} on a 7-point scale, 1 being strongly disagree to 7 being strongly agree. The total score was calculated by combining scores from all eight items, except for one item that was judged as unsuitable for the current experiment. Sample items included ``I strive to feel empathy for others.'' and `` It is good to feel empathy for others.'' 

Figure ~\ref{fig:2} shows the difference in mean ratings of empathic motives between the two groups. 
Participants in the combined group (\textit{M} = 40.59, \textit{SD} = 4.43) showed higher empathic motives scores than those in the VR-only group (\textit{M} = 37.59, \textit{SD} = 4.81). The difference between the two groups was significant (\textit{t}(52) = 2.38, \textit{p} = .02, \textit{d} = .65). In other words, participants who embraced a growth mindset via the article as well as VR experience had higher motivations to empathize with others than those who did not. 

\subsubsection{Empathy} Participants responded to Interpersonal Reactivity Index \cite{davis1980multidimensional, davis1983measuring} on a 5-point scale, 1 being strongly disagree to 5 being strongly agree. The empathy level was calculated by combining scores of 14 items, including ``When I'm upset at someone, I usually try to put myself in his shoes.''  According to the result of an independent t-test, the difference in the empathy scores between the combined group (\textit{M} = 49.44, \textit{SD} = 6.08) and the VR-only group (\textit{M} = 49.30, \textit{SD} = 5.89) was not statistically significant (\textit{t}(52) = .09, \textit{p} = .93, \textit{d} = .02). 

\subsubsection{Hypothetical Prosocial Behavior} Participants were asked how much money they would donate to charity for children if they got a bonus of KRW 20,000 (equivalent to US \$20) in return for their participation \cite{carlson2022belief, rathje2021attending}. According to the Mann-Whitney U Test \cite{mann1947test}, the difference in prosocial behavior between the combined group (\textit{M} = 15296, \textit{SD} = 5376) and the VR-only group (\textit{M} = 12370, \textit{SD} = 7540) was not statistically significant (\textit{W} = 432.5, \textit{p} = .22, \textit{d} = .17). This result means that the amounts of money participants decided to donate to charity did not have any differences between the two groups.

\subsection{Post-experiment Interview}
\subsubsection{Satisfaction with the Content} Participants usually thought that the sequence was apparent and the playing procedure was highly intuitive. \textit{Participant A} said, ``As the missions were diverse with boxes, cats, and traffic lights, I could continuously concentrate on the content. The difficulty of each mission was also appropriate, which made me smoothly play.'' \textit{Participant B} left the comment, ``The character setting and scenario of the superhero having to solve crises on the road were interesting for me.'' \textit{Participant C} reviewed, ``It was easy to understand how to play the content with the hand controllers. The sequence was also clear-cut.'' \textit{Participant D} liked the instructions in the content that explained who the virtual character was or how to manipulate the controllers. 

\subsubsection{Educational Effects and Mindsets} Participants also agreed with the educational effects of the content in promoting empathy. \textit{Participant A} commented, ``I thought `Yes, I am helping the NPCs right now!' while playing, and it seemed the VR content had educational effects.'' \textit{Participant B} said, ``I felt like I must help truck drivers, cats, or other citizens, and NPCs or sound effects made me think that way.'' \textit{Participant C} said, ``Helping others became familiar by playing the content. I felt like I could help others in the physical world based on this experience.'' \textit{Participant D} said that the content's background (e.g., city and roads) was similar to the actual world, instigating her to help others if similar situations actually happened.

\section{Discussion}
\subsection{Design Considerations} 
The goal of our VR content was to promote players' growth mindset and empathy through identification. Therefore, the hero character was created to bridge the gap between players' original and prosocial selves \cite{gupta2020investigating}. We also raised players' comfortableness by increasing the character's familiarity \cite{beeney2011feel, paiva2005learning}, making them smoothly identify with it and switch identities. The content's sequences were also designed to reinforce players' growth mindsets step by step. We predicted that their growth mindsets made them feel a sense of achievement and practice helping behaviors with effort \cite{dweck2017journey, mrazek2018expanding}. 

Additional factors were also considered to increase players' motivation, immersion, and achievement \cite{grant2010little, guo2017influence, mccullough2001gratitude, lieberman2006can}: the interactions with NPCs, the content's difficulty and duration, the light sources and colors of the background, the motion speed of dynamic objects, the interactive behaviors, and diverse sensory feedback.

\subsection{Effects on Promoting Empathy}
The current study conducted experiments to investigate whether the VR experience and mindset training fostered players' growth mindsets and empathy. We, therefore, compared the scores of empathic motives, empathy, and prosocial behavior between the combined and VR-only groups.

According to the results, participants in the combined group reported significantly higher motivations to empathize with others than participants in the VR-only group. The combined group also showed higher empathy and prosocial behavior levels, although the differences were insignificant. The results replicated those of the prior studies that embracing growth mindsets and practicing altruistic behaviors via training is an effective intervention \cite{weisz2021building, weisz2022brief, herrera2018building}, while suggesting a new intervention approach: the combination of the mindset training and VR experience. The participants' post-experiment interview responses repeatedly supported the effects. They answered that they were motivated to assist NPCs by completing the missions in the content. 

Empathy embraces a motivational aspect representing concern for others' emotional distress \cite{zaki2019war}. Therefore, stimulating empathic motives is essential in boosting empathy and prosocial behavior \cite{weisz2021building, weisz2022brief}. In this regard, the present study shows the possibility of increasing empathy and prosocial behaviors by modulating mindsets within VR-based interventions. 

\subsection{Practical Implications}
The present study suggests how Virtual Reality can be developed into a VR-based empathy intervention program for college students. VR provides immersive interactive situations, which enables identification with virtual characters and eventually changes players' behaviors \cite{gupta2020investigating, herrera2018building, schutte2017facilitating}. \textit{Our Neighbor Hero}, the VR content we implemented in this study, also gave players unique experiences: being a superhero and addressing neighborhood emergency issues with superpowers. These experiences led participants to embrace growth mindsets and higher empathic motives. The study indicates that empathy education using prosocial VR content can alter individuals' thoughts and attitudes by modulating mindsets.

The current study also investigates how VR experiences could resolve loneliness and intergroup conflicts. Previous research has shown that engaging in altruistic behaviors can reduce loneliness and improve the sense of fulfillment \cite{dossey2018helper}. By using the VR content in this study, lonely individuals can develop empathy, which increases social interactions and alleviates their feelings of isolation. Moreover, VR can be used as a powerful tool to see other people from different perspectives\footnote{for this reason, it's often referred to as `the empathy machine' \cite{milk2015}.}. Perspective-taking is a crucial aspect of empathic understanding, and this, in turn, can lead to empathic and effective communication \cite{de2006empathy, klimecki2019role}. This effect of VR on perspective-taking can also expand to out-group members \cite{halperin2011promoting}.

\subsection{Limitations and Future Work}

In the current study, we recruited only college students, similar to previous studies on the same subject \cite{weisz2021building}. Future studies should include more diverse participants, such as middle or high school students, as mindset-based empathy interventions have been shown to be effective in this age group \cite{weisz2022brief}. 
Future studies can also expand VR design to test empathic ability in different contexts that require users to practice prosocial behaviors. For example, the superhero character can escort their elderly neighbors or carry heavy things for them in a residential area. Otherwise, school or campus can be the main background where the character interacts with his schoolmates, considering the content was developed to increase college students' empathy and cooperation skills. Furthermore, putting players into others' shoes who are in a difficult situation (e.g., people without homes \cite{herrera2018building} or living in a refugee camp \cite{schutte2017facilitating}) can be another future scenario. 
    
\section{Conclusions}
In the current study, we investigated whether having a growth mindset of empathy would enhance individuals' empathy when practiced through VR content. For this reason, we developed our own VR content, \textit{Our Neighbor Hero}.  
We devised a concept of `superhero' in designing our VR content, inducing participants to engage in situations that call for helping others. The virtual superhero character was also designed for practicing a growth mindset of empathy; to achieve this, we designed the content sequence so that the virtual superhero develops its prosocial skills as the sequence progresses.
We found that participants reported higher empathic motives when they received both growth mindset training and experienced the VR content. Our findings suggest that exercising prosocial behaviors through VR content could effectively enhance the sense of community and aid in resolving social conflicts when paired with additional training on how empathy can be improved with practice. 

\begin{acks} 
This work was supported by the Seoul National University Faculty of Liberal Education. We thank Dohee Kim, Yong-Ha Park, and members of the Human Factors Psychology Lab for assistance with VR development.
\end{acks}

\bibliographystyle{ACM-Reference-Format}
\bibliography{superhero-bibl}


\begin{thebibliography}{59}


\ifx \showCODEN    \undefined \def \showCODEN     #1{\unskip}     \fi
\ifx \showDOI      \undefined \def \showDOI       #1{#1}\fi
\ifx \showISBNx    \undefined \def \showISBNx     #1{\unskip}     \fi
\ifx \showISBNxiii \undefined \def \showISBNxiii  #1{\unskip}     \fi
\ifx \showISSN     \undefined \def \showISSN      #1{\unskip}     \fi
\ifx \showLCCN     \undefined \def \showLCCN      #1{\unskip}     \fi
\ifx \shownote     \undefined \def \shownote      #1{#1}          \fi
\ifx \showarticletitle \undefined \def \showarticletitle #1{#1}   \fi
\ifx \showURL      \undefined \def \showURL       {\relax}        \fi
\providecommand\bibfield[2]{#2}
\providecommand\bibinfo[2]{#2}
\providecommand\natexlab[1]{#1}
\providecommand\showeprint[2][]{arXiv:#2}

\bibitem[Bailenson(2018)]%
        {bailenson2018experience}
\bibfield{author}{\bibinfo{person}{Jeremy Bailenson}.} \bibinfo{year}{2018}\natexlab{}.
\newblock \bibinfo{booktitle}{\emph{Experience on Demand: What Virtual Reality Is, How It Works, and What It Can Do}}.
\newblock \bibinfo{publisher}{WW Norton \& Company}, \bibinfo{address}{New York, NY}.
\newblock


\bibitem[Bailey et~al\mbox{.}(2016)]%
        {bailey2016does}
\bibfield{author}{\bibinfo{person}{Jakki~O Bailey}, \bibinfo{person}{Jeremy~N. Bailenson}, {and} \bibinfo{person}{Daniel Casasanto}.} \bibinfo{year}{2016}\natexlab{}.
\newblock \showarticletitle{When does virtual embodiment change our minds?}
\newblock \bibinfo{journal}{\emph{Presence: Teleoperators and Virtual Environments}} \bibinfo{volume}{25}, \bibinfo{number}{3} (\bibinfo{date}{Dec.} \bibinfo{year}{2016}), \bibinfo{pages}{222--233}.
\newblock
\urldef\tempurl%
\url{https://doi.org/10.1162/pres_a_00263}
\showDOI{\tempurl}


\bibitem[Batt-Rawden et~al\mbox{.}(2013)]%
        {batt2013teaching}
\bibfield{author}{\bibinfo{person}{Samantha~A. Batt-Rawden}, \bibinfo{person}{Margaret~S. Chisolm}, \bibinfo{person}{Blair Anton}, {and} \bibinfo{person}{Tabor~E. Flickinger}.} \bibinfo{year}{2013}\natexlab{}.
\newblock \showarticletitle{Teaching empathy to medical students: An updated, systematic review}.
\newblock \bibinfo{journal}{\emph{Academic Medicine}} \bibinfo{volume}{88}, \bibinfo{number}{8} (\bibinfo{date}{Aug.} \bibinfo{year}{2013}), \bibinfo{pages}{1171--1177}.
\newblock
\urldef\tempurl%
\url{https://doi.org/10.1097/acm.0b013e318299f3e3}
\showDOI{\tempurl}


\bibitem[Beeney et~al\mbox{.}(2011)]%
        {beeney2011feel}
\bibfield{author}{\bibinfo{person}{Joseph~E. Beeney}, \bibinfo{person}{Robert~G. Franklin~Jr.}, \bibinfo{person}{Kenneth~N. Levy}, {and} \bibinfo{person}{Reginald~B. Adams~Jr.}} \bibinfo{year}{2011}\natexlab{}.
\newblock \showarticletitle{I feel your pain: Emotional closeness modulates neural responses to empathically experienced rejection}.
\newblock \bibinfo{journal}{\emph{Social Neuroscience}} \bibinfo{volume}{6}, \bibinfo{number}{4} (\bibinfo{date}{Dec.} \bibinfo{year}{2011}), \bibinfo{pages}{369--376}.
\newblock
\urldef\tempurl%
\url{https://doi.org/10.1080/17470919.2011.557245}
\showDOI{\tempurl}


\bibitem[Buckley and Anderson(2006)]%
        {buckley2006theoretical}
\bibfield{author}{\bibinfo{person}{Katharine~E. Buckley} {and} \bibinfo{person}{Craig Anderson}.} \bibinfo{year}{2006}\natexlab{}.
\newblock \showarticletitle{A theoretical model of the effects and consequences of playing video games}.
\newblock In \bibinfo{booktitle}{\emph{Playing Video Games: Motives, Responses, and Consequences}}, \bibfield{editor}{\bibinfo{person}{P.~Vorderer} {and} \bibinfo{person}{J.~Bryant}} (Eds.). \bibinfo{publisher}{Lawrence Erlbaum Associates Publishers}, \bibinfo{address}{Mahwah, NJ}, \bibinfo{pages}{363--378}.
\newblock


\bibitem[Burke(2006)]%
        {burke2006identity}
\bibfield{author}{\bibinfo{person}{Peter~J. Burke}.} \bibinfo{year}{2006}\natexlab{}.
\newblock \showarticletitle{Identity change}.
\newblock \bibinfo{journal}{\emph{Social Psychology Quarterly}} \bibinfo{volume}{69}, \bibinfo{number}{1} (\bibinfo{date}{March} \bibinfo{year}{2006}), \bibinfo{pages}{81--96}.
\newblock
\urldef\tempurl%
\url{https://doi.org/10.1177/019027250606900106}
\showDOI{\tempurl}


\bibitem[Carlson and Zaki(2022)]%
        {carlson2022belief}
\bibfield{author}{\bibinfo{person}{Ryan~W. Carlson} {and} \bibinfo{person}{Jamil Zaki}.} \bibinfo{year}{2022}\natexlab{}.
\newblock \showarticletitle{Belief in altruistic motives predicts prosocial actions and inferences}.
\newblock \bibinfo{journal}{\emph{Psychological Reports}} \bibinfo{volume}{125}, \bibinfo{number}{4} (\bibinfo{date}{Aug.} \bibinfo{year}{2022}), \bibinfo{pages}{2191--2212}.
\newblock
\urldef\tempurl%
\url{https://doi.org/10.1177/00332941211013529}
\showDOI{\tempurl}


\bibitem[Davis(1980)]%
        {davis1980multidimensional}
\bibfield{author}{\bibinfo{person}{Mark~H. Davis}.} \bibinfo{year}{1980}\natexlab{}.
\newblock \showarticletitle{A multidimensional approach to individual differences in empathy}.
\newblock \bibinfo{journal}{\emph{JSAS Catalog of Selected Documents in Psychology}}  \bibinfo{volume}{10} (\bibinfo{year}{1980}), \bibinfo{pages}{85}.
\newblock


\bibitem[Davis(1983)]%
        {davis1983measuring}
\bibfield{author}{\bibinfo{person}{Mark~H. Davis}.} \bibinfo{year}{1983}\natexlab{}.
\newblock \showarticletitle{Measuring individual differences in empathy: Evidence for a multidimensional approach.}
\newblock \bibinfo{journal}{\emph{Journal of Personality and Social Psychology}} \bibinfo{volume}{44}, \bibinfo{number}{1} (\bibinfo{date}{Jan.} \bibinfo{year}{1983}), \bibinfo{pages}{113--126}.
\newblock
\urldef\tempurl%
\url{https://doi.org/10.1037/0022-3514.44.1.113}
\showDOI{\tempurl}


\bibitem[De~Vignemont and Singer(2006)]%
        {de2006vignemont}
\bibfield{author}{\bibinfo{person}{Frederique De~Vignemont} {and} \bibinfo{person}{Tania Singer}.} \bibinfo{year}{2006}\natexlab{}.
\newblock \showarticletitle{The empathic brain: how, when and why?}
\newblock \bibinfo{journal}{\emph{Trends in Cognitive Sciences}} \bibinfo{volume}{10}, \bibinfo{number}{10} (\bibinfo{date}{Oct.} \bibinfo{year}{2006}), \bibinfo{pages}{435--441}.
\newblock
\urldef\tempurl%
\url{https://doi.org/10.1016/j.tics.2006.08.008}
\showDOI{\tempurl}


\bibitem[De~Wied et~al\mbox{.}(2007)]%
        {de2006empathy}
\bibfield{author}{\bibinfo{person}{Minet De~Wied}, \bibinfo{person}{Susan J.~T. Branje}, {and} \bibinfo{person}{Wim H.~J. Meeus}.} \bibinfo{year}{2007}\natexlab{}.
\newblock \showarticletitle{Empathy and conflict resolution in friendship relations among adolescents}.
\newblock \bibinfo{journal}{\emph{Aggressive Behavior}}  \bibinfo{volume}{33} (\bibinfo{year}{2007}), \bibinfo{pages}{48--55}.
\newblock
\urldef\tempurl%
\url{https://doi.org/10.1002/ab.20166}
\showDOI{\tempurl}


\bibitem[Dossey(2018)]%
        {dossey2018helper}
\bibfield{author}{\bibinfo{person}{Larry Dossey}.} \bibinfo{year}{2018}\natexlab{}.
\newblock \showarticletitle{The helper's high}.
\newblock \bibinfo{journal}{\emph{Explore}} \bibinfo{volume}{14}, \bibinfo{number}{6} (\bibinfo{date}{Nov.} \bibinfo{year}{2018}), \bibinfo{pages}{393--399}.
\newblock
\urldef\tempurl%
\url{https://doi.org/10.1016/j.explore.2018.10.003}
\showDOI{\tempurl}


\bibitem[Dweck(1999)]%
        {dweck1999self}
\bibfield{author}{\bibinfo{person}{Carol~S. Dweck}.} \bibinfo{year}{1999}\natexlab{}.
\newblock \bibinfo{booktitle}{\emph{Self-Theories: Their Role in Motivation, Personality, and Development}}.
\newblock \bibinfo{publisher}{Psychology Press}, \bibinfo{address}{New York, NY}.
\newblock


\bibitem[Dweck(2006)]%
        {dweck2006mindset}
\bibfield{author}{\bibinfo{person}{Carol~S. Dweck}.} \bibinfo{year}{2006}\natexlab{}.
\newblock \bibinfo{booktitle}{\emph{Mindset: The New Psychology of Success}}.
\newblock \bibinfo{publisher}{Random House}, \bibinfo{address}{New York, NY}.
\newblock


\bibitem[Dweck(2012)]%
        {dweck2012mindsets}
\bibfield{author}{\bibinfo{person}{Carol~S. Dweck}.} \bibinfo{year}{2012}\natexlab{}.
\newblock \showarticletitle{Mindsets and human nature: Promoting change in the Middle East, the schoolyard, the racial divide, and willpower.}
\newblock \bibinfo{journal}{\emph{American Psychologist}} \bibinfo{volume}{67}, \bibinfo{number}{8} (\bibinfo{date}{Nov.} \bibinfo{year}{2012}), \bibinfo{pages}{614--622}.
\newblock
\urldef\tempurl%
\url{https://doi.org/10.1037/a0029783}
\showDOI{\tempurl}


\bibitem[Dweck(2017)]%
        {dweck2017journey}
\bibfield{author}{\bibinfo{person}{Carol~S. Dweck}.} \bibinfo{year}{2017}\natexlab{}.
\newblock \showarticletitle{The journey to children's mindsets—and beyond}.
\newblock \bibinfo{journal}{\emph{Child Development Perspectives}} \bibinfo{volume}{11}, \bibinfo{number}{2} (\bibinfo{date}{Jan.} \bibinfo{year}{2017}), \bibinfo{pages}{139--144}.
\newblock
\urldef\tempurl%
\url{https://doi.org/10.1111/cdep.12225}
\showDOI{\tempurl}


\bibitem[Dweck and Leggett(1988)]%
        {dweck1988social}
\bibfield{author}{\bibinfo{person}{Carol~S. Dweck} {and} \bibinfo{person}{Ellen~L. Leggett}.} \bibinfo{year}{1988}\natexlab{}.
\newblock \showarticletitle{A social-cognitive approach to motivation and personality.}
\newblock \bibinfo{journal}{\emph{Psychological Review}} \bibinfo{volume}{95}, \bibinfo{number}{2} (\bibinfo{date}{April} \bibinfo{year}{1988}), \bibinfo{pages}{256--273}.
\newblock
\urldef\tempurl%
\url{https://doi.org/10.1037/0033-295X.95.2.256}
\showDOI{\tempurl}


\bibitem[Eisenberg(2014)]%
        {eisenberg2014altruistic}
\bibfield{author}{\bibinfo{person}{Nancy Eisenberg}.} \bibinfo{year}{2014}\natexlab{}.
\newblock \bibinfo{booktitle}{\emph{Altruistic Emotion, Cognition, and Behavior (PLE: Emotion)}}.
\newblock \bibinfo{publisher}{Psychology Press}, \bibinfo{address}{New York, NY}.
\newblock


\bibitem[Eisenberg and Miller(1987)]%
        {eisenberg1987relation}
\bibfield{author}{\bibinfo{person}{Nancy Eisenberg} {and} \bibinfo{person}{Paul~A. Miller}.} \bibinfo{year}{1987}\natexlab{}.
\newblock \showarticletitle{The relation of empathy to prosocial and related behaviors.}
\newblock \bibinfo{journal}{\emph{Psychological Bulletin}} \bibinfo{volume}{101}, \bibinfo{number}{1} (\bibinfo{date}{Jan.} \bibinfo{year}{1987}), \bibinfo{pages}{91--119}.
\newblock
\urldef\tempurl%
\url{https://doi.org/10.1037/0033-2909.101.1.91}
\showDOI{\tempurl}


\bibitem[Eysenck and Eysenck(1978)]%
        {eysenck1978impulsiveness}
\bibfield{author}{\bibinfo{person}{Sybil B.~G. Eysenck} {and} \bibinfo{person}{Hans~J. Eysenck}.} \bibinfo{year}{1978}\natexlab{}.
\newblock \showarticletitle{Impulsiveness and venturesomeness: Their position in a dimensional system of personality description}.
\newblock \bibinfo{journal}{\emph{Psychological Reports}} \bibinfo{volume}{43}, \bibinfo{number}{3\_suppl} (\bibinfo{date}{Dec.} \bibinfo{year}{1978}), \bibinfo{pages}{1247--1255}.
\newblock
\urldef\tempurl%
\url{https://doi.org/10.2466/pr0.1978.43.3f.1247}
\showDOI{\tempurl}


\bibitem[Gentile et~al\mbox{.}(2009)]%
        {gentile2009effects}
\bibfield{author}{\bibinfo{person}{Douglas~A. Gentile}, \bibinfo{person}{Craig~A. Anderson}, \bibinfo{person}{Shintaro Yukawa}, \bibinfo{person}{Nobuko Ihori}, \bibinfo{person}{Muniba Saleem}, \bibinfo{person}{Lim~Kam Ming}, \bibinfo{person}{Akiko Shibuya}, \bibinfo{person}{Albert~K. Liau}, \bibinfo{person}{Angeline Khoo}, \bibinfo{person}{Brad~J. Bushman}, \bibinfo{person}{L.~Rowell Huesmann}, {and} \bibinfo{person}{Akira Sakamoto}.} \bibinfo{year}{2009}\natexlab{}.
\newblock \showarticletitle{The effects of prosocial video games on prosocial behaviors: International evidence from correlational, longitudinal, and experimental studies}.
\newblock \bibinfo{journal}{\emph{Personality and Social Psychology Bulletin}} \bibinfo{volume}{35}, \bibinfo{number}{6} (\bibinfo{date}{June} \bibinfo{year}{2009}), \bibinfo{pages}{752--763}.
\newblock
\urldef\tempurl%
\url{https://doi.org/10.1177/0146167209333045}
\showDOI{\tempurl}


\bibitem[Grant and Gino(2010)]%
        {grant2010little}
\bibfield{author}{\bibinfo{person}{Adam~M. Grant} {and} \bibinfo{person}{Francesca Gino}.} \bibinfo{year}{2010}\natexlab{}.
\newblock \showarticletitle{A little thanks goes a long way: Explaining why gratitude expressions motivate prosocial behavior.}
\newblock \bibinfo{journal}{\emph{Journal of Personality and Social Psychology}} \bibinfo{volume}{98}, \bibinfo{number}{6} (\bibinfo{date}{June} \bibinfo{year}{2010}), \bibinfo{pages}{946--955}.
\newblock
\urldef\tempurl%
\url{https://doi.org/10.1037/a0017935}
\showDOI{\tempurl}


\bibitem[Guo(2017)]%
        {guo2017influence}
\bibfield{author}{\bibinfo{person}{Yuan Guo}.} \bibinfo{year}{2017}\natexlab{}.
\newblock \showarticletitle{The Influence of Social Support on the Prosocial Behavior of College Students: The Mediating Effect Based on Interpersonal Trust.}
\newblock \bibinfo{journal}{\emph{English Language Teaching}} \bibinfo{volume}{10}, \bibinfo{number}{12} (\bibinfo{date}{Nov.} \bibinfo{year}{2017}), \bibinfo{pages}{158--163}.
\newblock
\urldef\tempurl%
\url{https://doi.org/10.5539/elt.v10n12p158}
\showDOI{\tempurl}


\bibitem[Gupta et~al\mbox{.}(2020)]%
        {gupta2020investigating}
\bibfield{author}{\bibinfo{person}{Saumya Gupta}, \bibinfo{person}{Theresa~Jean Tanenbaum}, \bibinfo{person}{Meena~Devii Muralikumar}, {and} \bibinfo{person}{Aparajita~S. Marathe}.} \bibinfo{year}{2020}\natexlab{}.
\newblock \showarticletitle{Investigating roleplaying and identity transformation in a virtual reality narrative experience}. In \bibinfo{booktitle}{\emph{Proceedings of the 2020 CHI Conference on Human Factors in Computing Systems (CHI '20)}}. \bibinfo{publisher}{ACM}, \bibinfo{address}{New York, NY}, \bibinfo{pages}{1--13}.
\newblock
\urldef\tempurl%
\url{https://doi.org/10.1145/3313831.3376762}
\showDOI{\tempurl}


\bibitem[Hallowell(1999)]%
        {hallowell1999human}
\bibfield{author}{\bibinfo{person}{Edward~M. Hallowell}.} \bibinfo{year}{1999}\natexlab{}.
\newblock \bibinfo{booktitle}{\emph{The Human Moment at Work}}.
\newblock \bibinfo{publisher}{Harvard Business Review}, \bibinfo{address}{Brighton, MA}.
\newblock


\bibitem[Halperin et~al\mbox{.}(2011)]%
        {halperin2011promoting}
\bibfield{author}{\bibinfo{person}{Eran Halperin}, \bibinfo{person}{Alexandra~G. Russell}, \bibinfo{person}{Kali~H. Trzesniewski}, \bibinfo{person}{James~J. Gross}, {and} \bibinfo{person}{Carol~S. Dweck}.} \bibinfo{year}{2011}\natexlab{}.
\newblock \showarticletitle{Promoting the Middle East peace process by changing beliefs about group malleability}.
\newblock \bibinfo{journal}{\emph{Science}} \bibinfo{volume}{333}, \bibinfo{number}{6050} (\bibinfo{date}{Aug.} \bibinfo{year}{2011}), \bibinfo{pages}{1767--1769}.
\newblock
\urldef\tempurl%
\url{https://doi.org/10.1126/science.1202925}
\showDOI{\tempurl}


\bibitem[Herrera et~al\mbox{.}(2018)]%
        {herrera2018building}
\bibfield{author}{\bibinfo{person}{Fernanda Herrera}, \bibinfo{person}{Jeremy Bailenson}, \bibinfo{person}{Erika Weisz}, \bibinfo{person}{Elise Ogle}, {and} \bibinfo{person}{Jamil Zaki}.} \bibinfo{year}{2018}\natexlab{}.
\newblock \showarticletitle{Building long-term empathy: A large-scale comparison of traditional and virtual reality perspective-taking}.
\newblock \bibinfo{journal}{\emph{PloS one}} \bibinfo{volume}{13}, \bibinfo{number}{10}, Article \bibinfo{articleno}{e0204494} (\bibinfo{date}{Oct.} \bibinfo{year}{2018}), \bibinfo{numpages}{37}~pages.
\newblock
\urldef\tempurl%
\url{https://doi.org/10.1371/journal.pone.0204494}
\showDOI{\tempurl}


\bibitem[Ho and Ng(2022)]%
        {ho2022perspective}
\bibfield{author}{\bibinfo{person}{Jeffrey C.~F. Ho} {and} \bibinfo{person}{Ryan Ng}.} \bibinfo{year}{2022}\natexlab{}.
\newblock \showarticletitle{Perspective-taking of non-player characters in prosocial virtual reality games: Effects on closeness, empathy, and game immersion}.
\newblock \bibinfo{journal}{\emph{Behaviour \& Information Technology}} \bibinfo{volume}{41}, \bibinfo{number}{6} (\bibinfo{date}{Dec.} \bibinfo{year}{2022}), \bibinfo{pages}{1185--1198}.
\newblock
\urldef\tempurl%
\url{https://doi.org/10.1080/0144929X.2020.1864018}
\showDOI{\tempurl}


\bibitem[Hofmeyer and Taylor(2021)]%
        {hofmeyer2021strategies}
\bibfield{author}{\bibinfo{person}{Anne Hofmeyer} {and} \bibinfo{person}{Ruth Taylor}.} \bibinfo{year}{2021}\natexlab{}.
\newblock \showarticletitle{Strategies and resources for nurse leaders to use to lead with empathy and prudence so they understand and address sources of anxiety among nurses practising in the era of COVID-19}.
\newblock \bibinfo{journal}{\emph{Journal of Clinical Nursing}} \bibinfo{volume}{30}, \bibinfo{number}{1-2} (\bibinfo{date}{Jan.} \bibinfo{year}{2021}), \bibinfo{pages}{298--305}.
\newblock
\urldef\tempurl%
\url{https://doi.org/10.1111/jocn.15520}
\showDOI{\tempurl}


\bibitem[Kelly and Barsade(2001)]%
        {kelly2001mood}
\bibfield{author}{\bibinfo{person}{Janice~R. Kelly} {and} \bibinfo{person}{Sigal~G. Barsade}.} \bibinfo{year}{2001}\natexlab{}.
\newblock \showarticletitle{Mood and emotions in small groups and work teams}.
\newblock \bibinfo{journal}{\emph{Organizational Behavior and Human Decision Processes}} \bibinfo{volume}{86}, \bibinfo{number}{1} (\bibinfo{date}{Sept.} \bibinfo{year}{2001}), \bibinfo{pages}{99--130}.
\newblock
\urldef\tempurl%
\url{https://doi.org/10.1006/obhd.2001.2974}
\showDOI{\tempurl}


\bibitem[Kelm et~al\mbox{.}(2014)]%
        {kelm2014interventions}
\bibfield{author}{\bibinfo{person}{Zak Kelm}, \bibinfo{person}{James Womer}, \bibinfo{person}{Jennifer~K Walter}, {and} \bibinfo{person}{Chris Feudtner}.} \bibinfo{year}{2014}\natexlab{}.
\newblock \showarticletitle{Interventions to cultivate physician empathy: a systematic review}.
\newblock \bibinfo{journal}{\emph{BMC Medical Education}} \bibinfo{volume}{14}, \bibinfo{number}{1}, Article \bibinfo{articleno}{219} (\bibinfo{date}{Oct.} \bibinfo{year}{2014}), \bibinfo{numpages}{11}~pages.
\newblock
\urldef\tempurl%
\url{https://doi.org/10.1186/1472-6920-14-219}
\showDOI{\tempurl}


\bibitem[Klimecki(2019)]%
        {klimecki2019role}
\bibfield{author}{\bibinfo{person}{Olga~M. Klimecki}.} \bibinfo{year}{2019}\natexlab{}.
\newblock \showarticletitle{The role of empathy and compassion in conflict resolution}.
\newblock \bibinfo{journal}{\emph{Emotion Review}} \bibinfo{volume}{11}, \bibinfo{number}{4} (\bibinfo{date}{Oct.} \bibinfo{year}{2019}), \bibinfo{pages}{310--325}.
\newblock
\urldef\tempurl%
\url{https://doi.org/10.1177/1754073919838609}
\showDOI{\tempurl}


\bibitem[Klimmt et~al\mbox{.}(2009)]%
        {klimmt2009video}
\bibfield{author}{\bibinfo{person}{Christoph Klimmt}, \bibinfo{person}{Doroth{\'e}e Hefner}, {and} \bibinfo{person}{Peter Vorderer}.} \bibinfo{year}{2009}\natexlab{}.
\newblock \showarticletitle{The video game experience as “true” identification: A theory of enjoyable alterations of players’ self-perception}.
\newblock \bibinfo{journal}{\emph{Communication Theory}} \bibinfo{volume}{19}, \bibinfo{number}{4} (\bibinfo{date}{Nov.} \bibinfo{year}{2009}), \bibinfo{pages}{351--373}.
\newblock
\urldef\tempurl%
\url{https://doi.org/10.1111/j.1468-2885.2009.01347.x}
\showDOI{\tempurl}


\bibitem[Lam et~al\mbox{.}(2011)]%
        {lam2011empathy}
\bibfield{author}{\bibinfo{person}{Tony Chiu~Ming Lam}, \bibinfo{person}{Klodiana Kolomitro}, {and} \bibinfo{person}{Flanny~C. Alamparambil}.} \bibinfo{year}{2011}\natexlab{}.
\newblock \showarticletitle{Empathy training: Methods, evaluation practices, and validity}.
\newblock \bibinfo{journal}{\emph{Journal of Multidisciplinary Evaluation}} \bibinfo{volume}{7}, \bibinfo{number}{16} (\bibinfo{date}{July} \bibinfo{year}{2011}), \bibinfo{pages}{162--200}.
\newblock
\urldef\tempurl%
\url{https://doi.org/10.56645/jmde.v7i16.314}
\showDOI{\tempurl}


\bibitem[Lieberman(2006)]%
        {lieberman2006can}
\bibfield{author}{\bibinfo{person}{Debra~A. Lieberman}.} \bibinfo{year}{2006}\natexlab{}.
\newblock \showarticletitle{What can we learn from playing interactive games?}
\newblock In \bibinfo{booktitle}{\emph{Playing Video Games: Motives, Responses, and Consequences}}, \bibfield{editor}{\bibinfo{person}{P.~Vorderer} {and} \bibinfo{person}{J.~Bryant}} (Eds.). \bibinfo{publisher}{Lawrence Erlbaum Associates Publishers}, \bibinfo{address}{Mahwah, NJ}, \bibinfo{pages}{447--469}.
\newblock


\bibitem[Mann and Whitney(1947)]%
        {mann1947test}
\bibfield{author}{\bibinfo{person}{Henry~B. Mann} {and} \bibinfo{person}{Donald~R. Whitney}.} \bibinfo{year}{1947}\natexlab{}.
\newblock \showarticletitle{On a test of whether one of two random variables is stochastically larger than the other}.
\newblock \bibinfo{journal}{\emph{The Annals of Mathematical Statistics}} \bibinfo{volume}{18}, \bibinfo{number}{1} (\bibinfo{date}{March} \bibinfo{year}{1947}), \bibinfo{pages}{50--60}.
\newblock
\urldef\tempurl%
\url{https://doi.org/10.1214/aoms/1177730491}
\showDOI{\tempurl}


\bibitem[McCullough et~al\mbox{.}(2001)]%
        {mccullough2001gratitude}
\bibfield{author}{\bibinfo{person}{Michael~E. McCullough}, \bibinfo{person}{Shelley~D. Kilpatrick}, \bibinfo{person}{Robert~A. Emmons}, {and} \bibinfo{person}{David~B. Larson}.} \bibinfo{year}{2001}\natexlab{}.
\newblock \showarticletitle{Is gratitude a moral affect?}
\newblock \bibinfo{journal}{\emph{Psychological Bulletin}} \bibinfo{volume}{127}, \bibinfo{number}{2} (\bibinfo{date}{March} \bibinfo{year}{2001}), \bibinfo{pages}{249--266}.
\newblock
\urldef\tempurl%
\url{https://doi.org/10.1037/0033-2909.127.2.249}
\showDOI{\tempurl}


\bibitem[Milk(2015)]%
        {milk2015}
\bibfield{author}{\bibinfo{person}{Chris Milk}.} \bibinfo{year}{2015}\natexlab{}.
\newblock \bibinfo{title}{How virtual reality can create the ultimate empathy machine}.
\newblock
\newblock
\urldef\tempurl%
\url{https://www.ted.com/talks/chris_milk_how_virtual_reality_can_create_the_ultimate_empathy_machine}
\showURL{%
Retrieved September 21, 2023 from \tempurl}


\bibitem[Mrazek et~al\mbox{.}(2018)]%
        {mrazek2018expanding}
\bibfield{author}{\bibinfo{person}{Alissa~J. Mrazek}, \bibinfo{person}{Elliott~D. Ihm}, \bibinfo{person}{Daniel~C. Molden}, \bibinfo{person}{Michael~D. Mrazek}, \bibinfo{person}{Claire~M. Zedelius}, {and} \bibinfo{person}{Jonathan~W. Schooler}.} \bibinfo{year}{2018}\natexlab{}.
\newblock \showarticletitle{Expanding minds: Growth mindsets of self-regulation and the influences on effort and perseverance}.
\newblock \bibinfo{journal}{\emph{Journal of Experimental Social Psychology}}  \bibinfo{volume}{79} (\bibinfo{date}{Nov.} \bibinfo{year}{2018}), \bibinfo{pages}{164--180}.
\newblock
\urldef\tempurl%
\url{https://doi.org/10.1016/j.jesp.2018.07.003}
\showDOI{\tempurl}


\bibitem[Neff(2003)]%
        {neff2003self}
\bibfield{author}{\bibinfo{person}{Kristin Neff}.} \bibinfo{year}{2003}\natexlab{}.
\newblock \showarticletitle{Self-compassion: An alternative conceptualization of a healthy attitude toward oneself}.
\newblock \bibinfo{journal}{\emph{Self and Identity}} \bibinfo{volume}{2}, \bibinfo{number}{2} (\bibinfo{year}{2003}), \bibinfo{pages}{85--101}.
\newblock
\urldef\tempurl%
\url{https://doi.org/10.1080/15298860309032}
\showDOI{\tempurl}


\bibitem[Paiva et~al\mbox{.}(2005)]%
        {paiva2005learning}
\bibfield{author}{\bibinfo{person}{Ana Paiva}, \bibinfo{person}{Jo{\~a}o Dias}, \bibinfo{person}{Daniel Sobral}, \bibinfo{person}{Ruth Aylett}, \bibinfo{person}{Sarah Woods}, \bibinfo{person}{Lynne Hall}, {and} \bibinfo{person}{Carsten Zoll}.} \bibinfo{year}{2005}\natexlab{}.
\newblock \showarticletitle{Learning by feeling: Evoking empathy with synthetic characters}.
\newblock \bibinfo{journal}{\emph{Applied Artificial Intelligence}} \bibinfo{volume}{19}, \bibinfo{number}{3-4} (\bibinfo{year}{2005}), \bibinfo{pages}{235--266}.
\newblock
\urldef\tempurl%
\url{https://doi.org/10.1080/08839510590910165}
\showDOI{\tempurl}


\bibitem[Paunesku et~al\mbox{.}(2015)]%
        {paunesku2015mind}
\bibfield{author}{\bibinfo{person}{David Paunesku}, \bibinfo{person}{Gregory~M. Walton}, \bibinfo{person}{Carissa Romero}, \bibinfo{person}{Eric~N. Smith}, \bibinfo{person}{David~S. Yeager}, {and} \bibinfo{person}{Carol~S. Dweck}.} \bibinfo{year}{2015}\natexlab{}.
\newblock \showarticletitle{Mind-set interventions are a scalable treatment for academic underachievement}.
\newblock \bibinfo{journal}{\emph{Psychological Science}} \bibinfo{volume}{26}, \bibinfo{number}{6} (\bibinfo{date}{June} \bibinfo{year}{2015}), \bibinfo{pages}{784--793}.
\newblock
\urldef\tempurl%
\url{https://doi.org/10.1177/0956797615571017}
\showDOI{\tempurl}


\bibitem[Pfattheicher et~al\mbox{.}(2022)]%
        {pfattheicher2022information}
\bibfield{author}{\bibinfo{person}{Stefan Pfattheicher}, \bibinfo{person}{Michael~Bang Petersen}, {and} \bibinfo{person}{Robert B{\"o}hm}.} \bibinfo{year}{2022}\natexlab{}.
\newblock \showarticletitle{Information about herd immunity through vaccination and empathy promote COVID-19 vaccination intentions.}
\newblock \bibinfo{journal}{\emph{Health Psychology}} \bibinfo{volume}{41}, \bibinfo{number}{2} (\bibinfo{date}{Feb.} \bibinfo{year}{2022}), \bibinfo{pages}{85--93}.
\newblock
\urldef\tempurl%
\url{https://doi.org/10.1037/hea0001096}
\showDOI{\tempurl}


\bibitem[Rathje et~al\mbox{.}(2021)]%
        {rathje2021attending}
\bibfield{author}{\bibinfo{person}{Steve Rathje}, \bibinfo{person}{Leor Hackel}, {and} \bibinfo{person}{Jamil Zaki}.} \bibinfo{year}{2021}\natexlab{}.
\newblock \showarticletitle{Attending live theatre improves empathy, changes attitudes, and leads to pro-social behavior}.
\newblock \bibinfo{journal}{\emph{Journal of Experimental Social Psychology}}  \bibinfo{volume}{95}, Article \bibinfo{articleno}{104138} (\bibinfo{date}{July} \bibinfo{year}{2021}), \bibinfo{numpages}{10}~pages.
\newblock
\urldef\tempurl%
\url{https://doi.org/10.1016/j.jesp.2021.104138}
\showDOI{\tempurl}


\bibitem[Sanchez-Vives and Slater(2005)]%
        {sanchez2005presence}
\bibfield{author}{\bibinfo{person}{Maria~V. Sanchez-Vives} {and} \bibinfo{person}{Mel Slater}.} \bibinfo{year}{2005}\natexlab{}.
\newblock \showarticletitle{From presence to consciousness through virtual reality}.
\newblock \bibinfo{journal}{\emph{Nature Reviews Neuroscience}}  \bibinfo{volume}{6} (\bibinfo{date}{April} \bibinfo{year}{2005}), \bibinfo{pages}{332--339}.
\newblock
\urldef\tempurl%
\url{https://doi.org/10.1038/nrn1651}
\showDOI{\tempurl}


\bibitem[Schumann et~al\mbox{.}(2014)]%
        {schumann2014addressing}
\bibfield{author}{\bibinfo{person}{Karina Schumann}, \bibinfo{person}{Jamil Zaki}, {and} \bibinfo{person}{Carol~S. Dweck}.} \bibinfo{year}{2014}\natexlab{}.
\newblock \showarticletitle{Addressing the empathy deficit: beliefs about the malleability of empathy predict effortful responses when empathy is challenging.}
\newblock \bibinfo{journal}{\emph{Journal of Personality and Social Psychology}} \bibinfo{volume}{107}, \bibinfo{number}{3} (\bibinfo{date}{Sept.} \bibinfo{year}{2014}), \bibinfo{pages}{475--493}.
\newblock
\urldef\tempurl%
\url{https://doi.org/10.1037/a0036738}
\showDOI{\tempurl}


\bibitem[Schutte and Stilinovi{\'c}(2017)]%
        {schutte2017facilitating}
\bibfield{author}{\bibinfo{person}{Nicola~S. Schutte} {and} \bibinfo{person}{Emma~J. Stilinovi{\'c}}.} \bibinfo{year}{2017}\natexlab{}.
\newblock \showarticletitle{Facilitating empathy through virtual reality}.
\newblock \bibinfo{journal}{\emph{Motivation and Emotion}}  \bibinfo{volume}{41} (\bibinfo{date}{Oct.} \bibinfo{year}{2017}), \bibinfo{pages}{708--712}.
\newblock
\urldef\tempurl%
\url{https://doi.org/10.1007/s11031-017-9641-7}
\showDOI{\tempurl}


\bibitem[Shams and Seitz(2008)]%
        {shams2008benefits}
\bibfield{author}{\bibinfo{person}{Ladan Shams} {and} \bibinfo{person}{Aaron~R. Seitz}.} \bibinfo{year}{2008}\natexlab{}.
\newblock \showarticletitle{Benefits of multisensory learning}.
\newblock \bibinfo{journal}{\emph{Trends in Cognitive Sciences}} \bibinfo{volume}{12}, \bibinfo{number}{11} (\bibinfo{date}{Sept.} \bibinfo{year}{2008}), \bibinfo{pages}{411--417}.
\newblock
\urldef\tempurl%
\url{https://doi.org/10.1016/j.tics.2008.07.006}
\showDOI{\tempurl}


\bibitem[Shapiro and Wilk(1965)]%
        {shapiro1965analysis}
\bibfield{author}{\bibinfo{person}{Samuel~S. Shapiro} {and} \bibinfo{person}{Martin~B. Wilk}.} \bibinfo{year}{1965}\natexlab{}.
\newblock \showarticletitle{An analysis of variance test for normality (complete samples)}.
\newblock \bibinfo{journal}{\emph{Biometrika}} \bibinfo{volume}{52}, \bibinfo{number}{3/4} (\bibinfo{date}{Dec.} \bibinfo{year}{1965}), \bibinfo{pages}{591--611}.
\newblock
\urldef\tempurl%
\url{https://doi.org/10.2307/2333709}
\showDOI{\tempurl}


\bibitem[Sigrist et~al\mbox{.}(2013)]%
        {sigrist2013augmented}
\bibfield{author}{\bibinfo{person}{Roland Sigrist}, \bibinfo{person}{Georg Rauter}, \bibinfo{person}{Robert Riener}, {and} \bibinfo{person}{Peter Wolf}.} \bibinfo{year}{2013}\natexlab{}.
\newblock \showarticletitle{Augmented visual, auditory, haptic, and multimodal feedback in motor learning: A review}.
\newblock \bibinfo{journal}{\emph{Psychonomic Bulletin \& Review}}  \bibinfo{volume}{20} (\bibinfo{date}{Nov.} \bibinfo{year}{2013}), \bibinfo{pages}{21--53}.
\newblock
\urldef\tempurl%
\url{https://doi.org/10.3758/s13423-012-0333-8}
\showDOI{\tempurl}


\bibitem[Teding~van Berkhout and Malouff(2016)]%
        {teding2016efficacy}
\bibfield{author}{\bibinfo{person}{Emily Teding~van Berkhout} {and} \bibinfo{person}{John~M. Malouff}.} \bibinfo{year}{2016}\natexlab{}.
\newblock \showarticletitle{The efficacy of empathy training: A meta-analysis of randomized controlled trials.}
\newblock \bibinfo{journal}{\emph{Journal of Counseling Psychology}} \bibinfo{volume}{63}, \bibinfo{number}{1} (\bibinfo{date}{July} \bibinfo{year}{2016}), \bibinfo{pages}{32--41}.
\newblock
\urldef\tempurl%
\url{https://doi.org/10.1037/cou0000093}
\showDOI{\tempurl}


\bibitem[Tusche et~al\mbox{.}(2016)]%
        {tusche2016decoding}
\bibfield{author}{\bibinfo{person}{Anita Tusche}, \bibinfo{person}{Anne B{\"o}ckler}, \bibinfo{person}{Philipp Kanske}, \bibinfo{person}{Fynn-Mathis Trautwein}, {and} \bibinfo{person}{Tania Singer}.} \bibinfo{year}{2016}\natexlab{}.
\newblock \showarticletitle{Decoding the charitable brain: Empathy, perspective taking, and attention shifts differentially predict altruistic giving}.
\newblock \bibinfo{journal}{\emph{Journal of Neuroscience}} \bibinfo{volume}{36}, \bibinfo{number}{17} (\bibinfo{date}{April} \bibinfo{year}{2016}), \bibinfo{pages}{4719--4732}.
\newblock
\urldef\tempurl%
\url{https://doi.org/10.1523/JNEUROSCI.3392-15.2016}
\showDOI{\tempurl}


\bibitem[Van~Loon et~al\mbox{.}(2018)]%
        {van2018virtual}
\bibfield{author}{\bibinfo{person}{Austin Van~Loon}, \bibinfo{person}{Jeremy Bailenson}, \bibinfo{person}{Jamil Zaki}, \bibinfo{person}{Joshua Bostick}, {and} \bibinfo{person}{Robb Willer}.} \bibinfo{year}{2018}\natexlab{}.
\newblock \showarticletitle{Virtual reality perspective-taking increases cognitive empathy for specific others}.
\newblock \bibinfo{journal}{\emph{PloS one}} \bibinfo{volume}{13}, \bibinfo{number}{8}, Article \bibinfo{articleno}{e0202442} (\bibinfo{date}{Aug.} \bibinfo{year}{2018}), \bibinfo{numpages}{19}~pages.
\newblock
\urldef\tempurl%
\url{https://doi.org/10.1371/journal.pone.0202442}
\showDOI{\tempurl}


\bibitem[Weber et~al\mbox{.}(2021)]%
        {weber2021get}
\bibfield{author}{\bibinfo{person}{Stefan Weber}, \bibinfo{person}{David Weibel}, {and} \bibinfo{person}{Fred~W. Mast}.} \bibinfo{year}{2021}\natexlab{}.
\newblock \showarticletitle{How to get there when you are there already? Defining presence in virtual reality and the importance of perceived realism}.
\newblock \bibinfo{journal}{\emph{Frontiers in Psychology}}  \bibinfo{volume}{12}, Article \bibinfo{articleno}{628298} (\bibinfo{date}{May} \bibinfo{year}{2021}), \bibinfo{numpages}{10}~pages.
\newblock
\urldef\tempurl%
\url{https://doi.org/10.3389/fpsyg.2021.628298}
\showDOI{\tempurl}


\bibitem[Weisz et~al\mbox{.}(2022)]%
        {weisz2022brief}
\bibfield{author}{\bibinfo{person}{Erika Weisz}, \bibinfo{person}{Patricia Chen}, \bibinfo{person}{Desmond~C. Ong}, \bibinfo{person}{Ryan~W. Carlson}, \bibinfo{person}{Marissa~D. Clark}, {and} \bibinfo{person}{Jamil Zaki}.} \bibinfo{year}{2022}\natexlab{}.
\newblock \showarticletitle{A brief intervention to motivate empathy among middle school students.}
\newblock \bibinfo{journal}{\emph{Journal of Experimental Psychology: General}} \bibinfo{volume}{151}, \bibinfo{number}{12} (\bibinfo{date}{Dec.} \bibinfo{year}{2022}), \bibinfo{pages}{3144--3153}.
\newblock
\urldef\tempurl%
\url{https://doi.org/10.1037/xge0001249}
\showDOI{\tempurl}


\bibitem[Weisz et~al\mbox{.}(2021)]%
        {weisz2021building}
\bibfield{author}{\bibinfo{person}{Erika Weisz}, \bibinfo{person}{Desmond~C. Ong}, \bibinfo{person}{Ryan~W. Carlson}, {and} \bibinfo{person}{Jamil Zaki}.} \bibinfo{year}{2021}\natexlab{}.
\newblock \showarticletitle{Building empathy through motivation-based interventions.}
\newblock \bibinfo{journal}{\emph{Emotion}} \bibinfo{volume}{21}, \bibinfo{number}{5} (\bibinfo{date}{Dec.} \bibinfo{year}{2021}), \bibinfo{pages}{990--999}.
\newblock
\urldef\tempurl%
\url{https://doi.org/10.1037/emo0000929}
\showDOI{\tempurl}


\bibitem[Yeager and Dweck(2012)]%
        {yeager2012mindsets}
\bibfield{author}{\bibinfo{person}{David~Scott Yeager} {and} \bibinfo{person}{Carol~S. Dweck}.} \bibinfo{year}{2012}\natexlab{}.
\newblock \showarticletitle{Mindsets that promote resilience: When students believe that personal characteristics can be developed}.
\newblock \bibinfo{journal}{\emph{Educational Psychologist}} \bibinfo{volume}{47}, \bibinfo{number}{4} (\bibinfo{date}{Oct.} \bibinfo{year}{2012}), \bibinfo{pages}{302--314}.
\newblock
\urldef\tempurl%
\url{https://doi.org/10.1080/00461520.2012.722805}
\showDOI{\tempurl}


\bibitem[Yeager et~al\mbox{.}(2019)]%
        {yeager2019national}
\bibfield{author}{\bibinfo{person}{David~S. Yeager}, \bibinfo{person}{Paul Hanselman}, \bibinfo{person}{Gregory~M. Walton}, \bibinfo{person}{Jared~S. Murray}, \bibinfo{person}{Robert Crosnoe}, \bibinfo{person}{Chandra Muller}, \bibinfo{person}{Elizabeth Tipton}, \bibinfo{person}{Barbara Schneider}, \bibinfo{person}{Chris~S. Hulleman}, \bibinfo{person}{Cintia~P. Hinojosa}, \bibinfo{person}{David Paunesku}, \bibinfo{person}{Carissa Romero}, \bibinfo{person}{Kate Flint}, \bibinfo{person}{Alice Roberts}, \bibinfo{person}{Jill Trott}, \bibinfo{person}{Ronaldo Iachan}, \bibinfo{person}{Jenny Buontempo}, \bibinfo{person}{Sophia~Man Yang}, \bibinfo{person}{Carlos~M. Carvalho}, \bibinfo{person}{P.~Richard Hahn}, \bibinfo{person}{Maithreyi Gopalan}, \bibinfo{person}{Pratik Mhatre}, \bibinfo{person}{Ronald Ferguson}, \bibinfo{person}{Angela~L. Duckworth}, {and} \bibinfo{person}{Carol~S. Dweck}.} \bibinfo{year}{2019}\natexlab{}.
\newblock \showarticletitle{A national experiment reveals where a growth mindset improves achievement}.
\newblock \bibinfo{journal}{\emph{Nature}} \bibinfo{volume}{573}, \bibinfo{number}{7774} (\bibinfo{date}{Aug.} \bibinfo{year}{2019}), \bibinfo{pages}{364--369}.
\newblock
\urldef\tempurl%
\url{https://doi.org/10.1038/s41586-019-1466-y}
\showDOI{\tempurl}


\bibitem[Zaki(2019)]%
        {zaki2019war}
\bibfield{author}{\bibinfo{person}{Jamil Zaki}.} \bibinfo{year}{2019}\natexlab{}.
\newblock \bibinfo{booktitle}{\emph{The War for Kindness: Building Empathy in a Fractured World}}.
\newblock \bibinfo{publisher}{Crown}, \bibinfo{address}{New York, NY}.
\newblock


\end{thebibliography}

\end{document}